\documentclass[prd,showpacs,amsmath,amssymb,nofootinbib,superscriptaddress]{revtex4}
\usepackage{array,mathtools,amssymb,booktabs,multirow,diagbox,hhline}
\pdfoutput=1

\usepackage{graphicx}
\usepackage{dsfont}
\usepackage{bbold}

\newcommand{\Od}{{\cal O}}
\newcommand{\diag}{\mbox{diag}}
\newcommand{\intT}{\int_0^{1/T} d\tau \int d^3 \vec{x}}

\newcommand{\mean}[1]{\left\langle{#1}\right\rangle}

\newcommand{\conds}{\langle \bar s s \rangle}
\newcommand{\condl}{\mean{\bar q q}_l}

\newcommand{\ID}{{\mathbb{1}}}

\newcommand{\eqchiral}{\!\stackrel{O(4)}{\sim}\!}
\newcommand{\equa}{\!\stackrel{U(1)_A}{\sim}\!}

\begin{document}

\title{Patterns and partners for chiral symmetry restoration}
\author{A. G\'omez Nicola}
\email{gomez@fis.ucm.es}
\affiliation{Departamento de F\'{\i}sica
Te\'orica and UPARCOS. Univ. Complutense. 28040 Madrid. Spain}
\author{J. Ruiz de Elvira}
\email{elvira@itp.unibe.ch}
\affiliation{Helmholtz-Institut f\"ur Strahlen- und Kernphysik, Universit\"at Bonn, D-53115 Bonn, Germany}
\affiliation{Albert Einstein Center for Fundamental Physics, Institute for Theoretical Physics,
University of Bern, Sidlerstrasse 5, CH--3012 Bern, Switzerland}

\begin{abstract}
We present and analyze a new set of Ward Identities which shed light on the distinction between different patterns of chiral symmetry restoration in QCD, namely $O(4)$ vs $O(4)\times U(1)_A$. 
The degeneracy of chiral partners for all scalar and pseudoscalar meson nonet members are studied through their corresponding correlators. 
 Around  chiral symmetry degeneration of $O(4)$ partners, our analysis predicts that $U(1)_A$ partners are  also degenerated. Our analysis also leads to  $I=1/2$ scalar-pseudoscalar partner degeneration at  exact chiral restoration and supports  ideal mixing between the $\eta$-$\eta^\prime$ and the $f_0(500)$-$f_0(980)$ mesons at $O(4)\times U(1)_A$ restoration, 
with a possible range where the pseudoscalar mixing vanishes if the two transitions are well separated.
We test our results with lattice data and provide further relevant observables regarding chiral and $U(1)_A$ restoration for future lattice and model analyses.
\end{abstract}

 
 \maketitle

\section{Introduction}
Chiral symmetry restoration is a prominent feature of the QCD phase diagram, realized in lattice simulations and presumably in matter formed after a Heavy Ion Collision.  
For vanishing baryon density and two massless flavors, a chiral restoring phase transition takes place with vanishing quark condensate and divergent scalar susceptibility, 
corresponding to $SU(2)_L\times SU(2)_R\sim O(4)$ restoration~\cite{Pisarski:1983ms,Smilga:1995qf}. 
For $N_f=2+1$ flavors and physical quark masses, $m_u=m_d=\hat m \ll m_s$, a crossover is expected  at a transition temperature $T_c\sim 155$ MeV~\cite{Aoki:2009sc,Bazavov:2011nk,Buchoff:2013nra},  
signaled by the inflection point of the light quark condensate $\condl$ and the peak of the scalar susceptibility.  
In addition, as we will detail below, another signal of chiral restoration would be the degeneration of chiral partners albeit, 
due to the crossover nature of the transition, different chiral restoring observables may lead to different transition temperatures. 
In the $\hat m/m_s\rightarrow 0^+$ limit (light chiral limit) all chiral restoration effects are enhanced, i.e, the quark condensate decreases, the scalar susceptibility peak increases~\cite{Ejiri:2009ac} and the degeneration of chiral partners becomes more noticeable, with  transition temperatures approaching the same value.  
In this work, we will use  the symbol $\eqchiral$ to mean equivalence under the $O(4)$ chiral group, which formally holds in the ideal regime of exact chiral restoration. 

In addition, the anomalous axial $U(1)_A$ symmetry can be asymptotically restored, driven by the vanishing of the instanton density~\cite{Gross:1980br}.
An ongoing debate is  then whether  $U(1)_A$  is  restored at the chiral transition.
If so, the restoration pattern would be $O(4)\times U(1)_A$ instead of $O(4)$ for two massless flavors and the order of the transition would change from second to first order~\cite{Pisarski:1983ms,Eser:2015pka}. The restoration of $U(1)_A$ also affects the chiral transition order for three flavors~\cite{Pelissetto:2013hqa}, 
as well as the behavior near the critical end point at finite temperature and baryon density~\cite{Mitter:2013fxa}.  It is important to emphasize that, unlike chiral restoration, which corresponds to a spontaneous symmetry breaking, the $U(1)_A$ is restored only asymptotically. Nevertheless, as we will discuss below, there are also $U(1)_A$ partners which become approximately degenerated above a certain temperature region. In that particular sense, we will use the symbol  $\equa$ to denote equivalence in such $U(1)_A$ restoration regime. 
    
The implications for the hadron spectrum are crucial.
The restoration of a global symmetry implies a degeneracy in the spectrum of particles, which is customarily studied through the behavior of their correlation functions. These correlators are  meant to be very sensitive to the transition from the ordered to the disordered state.
The hadronic states becoming degenerate at chiral restoration are usually known as chiral partners.  In more detail, the pion is expected to degenerate with the $\sigma/f_0(500)$ meson  within a $O(4)$ pattern, 
whereas the restoration of the $U(1)_A$ symmetry would also degenerate the pion and the $a_0(980)$, i.e. the member of the scalar nonet with the same pion quantum numbers but an opposite parity.
It is also natural to investigate the fate of the rest of the members of the scalar and pseudoscalar nonet, 
i.e the $\kappa$/$K_0(800)$ versus the kaon for $I=1/2$, and the $f_0(980)-f_0(500)$ pair versus the $\eta-\eta'$ for $I=0$, which has not been done before in the present context.  
Note that the breaking of Lorentz covariance in the thermal bath must be taken into account when defining the spectral properties of those hadrons through their correlation functions~\cite{Kapusta:2006pm}.

Moreover, if chiral and $U(1)_A$ restoration happen to be close, a proper description of light meson phenomenology at finite temperature will require the inclusion of the $\eta'$ as the ninth Goldstone boson, as in the large-$N_c$ framework~\cite{wittenetal}.  
In fact, there is experimental evidence of the reduction of the $\eta'$ mass in the hot medium~\cite{Csorgo:2009pa}, pointing out to $U(1)_A$ restoration.
This should result in an increase of the $\eta^\prime$ production cross section, which might be observed in dilepton and diphoton experiments at finite temperature~\cite{Kapusta:1995ww}. The reduction of the $\eta'$ mass in the nuclear medium and its connection with chiral symmetry restoration has been also analyzed in~\cite{Jido:2011pq}.
  
The idea that $U(1)_A$ partners can degenerate in an ideal chiral restoring scenario was suggested in~\cite {Shuryak:1993ee} and corroborated in~\cite{Cohen:1996ng} through an analysis of spectral properties of the QCD quark propagator. Nevertheless, in the real world with massive quarks, nontrivial gauge configurations make in general a nonzero $U(1)_A$ breaking to be present~\cite{Lee:1996zy}, even though $U(1)_A$ partners could be approximately degenerate.
Later effective models and renormalization-group approaches to this problem can be found in~\cite{Eser:2015pka,Pelissetto:2013hqa,Mitter:2013fxa,Meggiolaro:2013swa,Fejos:2015xca,Heller:2015box,Ishii:2015ira,Ishii:2016dln}. 

Chiral partners and patterns have  also been  recently examined by different lattice collaborations. 
Nevertheless, there is currently no consensus on the restoration scenario.
On the one hand, a $O(4)$ pattern has been proposed in~\cite{Buchoff:2013nra}, with nonzero quark masses and $N_f=2+1$. 
Namely, in that work $\pi-a_0$ and other  $U(1)_A$ symmetry partners degenerate asymptotically but their difference is still sizable near the point where $\pi-\sigma$ degeneracy occurs.
On the other hand, a $O(4)\times U(1)_A$ pattern for $N_f=2$ has been suggested in~\cite{Cossu:2013uua,Tomiya:2016jwr,Aoki:2012yj} near the chiral limit and in~\cite{Brandt:2016daq} for the massive case, the latter through the analysis of screening masses. 
Parity degeneracy in the baryon sector has also been studied in~\cite{Aarts}. 
In this context, it is important to mention that lattice measurements involving $U(1)_A$-related correlators require great care, due to the sampling of the different topological sectors~\cite{Borsanyi:2016ksw,Petreczky:2016vrs}.

Our aim in this work is to provide new results to shed light on chiral patterns and partner degeneracy. For that sake, we will rely on  Ward Identities (WI) derived formally in QCD with $N_f=2+1$ flavors. 
Since, by definition, the WI construction is model-independent, our results  could be tested in lattice and model analyses. In addition, the relation between different WI will help to  understand the current controversy about the symmetry breaking pattern. 
This paper is organized as follows. In section~\ref{sec:wi} the relevant WI are derived, while their various consequences for chiral and $U(1)_A$ restoration of the full meson nonet are analyzed in detail in section~\ref{sec:cons}.  
Our main conclusions are summarized in section~\ref{sec:conc}. 
 
\section{Ward identities}
\label{sec:wi}
We will start considering an infinitesimal vector and axial transformation on a quark field $\psi'=\psi+\delta\psi$,
\begin{equation}
\delta\psi(x)=i\left(\alpha_V^a(x)\frac{ \lambda_a}{2}+\alpha_A^a(x)\frac{ \lambda_a}{2}\gamma_5\right)\psi(x),\nonumber
\end{equation}
with $\lambda^{a=1,\dots 8}$ the $SU(3)$ Gell-Mann matrices and $\lambda^0=\sqrt{2/3}\,\mathds{1}$. 
As explained in~\cite{Nicola:2016jlj},  the QCD expectation value of a pseudoscalar operator $\mathcal{O}$ in terms of the transformed fields leads to
\begin{align}
\left\langle\frac{\delta\mathcal{O}(y)}{\delta \alpha_A^a(x)}\right\rangle=&
-\left\langle\mathcal{O}(y)\bar\psi(x)\left\{\frac{\lambda^a}{2},\mathcal{M}\right\}\gamma_5\psi(x)\right\rangle+i\frac{\delta_{a0}}{\sqrt{6}}\left\langle \mathcal{O}(y) A(x)\right\rangle,\label{wigen}\\
&\nonumber\\
\left\langle\frac{\delta\mathcal{O}(y)}{\delta \alpha_V^a(x)}\right\rangle=&\left\langle\mathcal{O}(y)\bar\psi(x)\left[\frac{\lambda^a}{2},\mathcal{M}\right]\psi(x)\right\rangle,
\label{wigenvec}
\end{align}
with $A(x)=\frac{3g^2}{16\pi^2}\mbox{Tr}_c G_{\mu\nu}\tilde G^{\mu\nu}$ the anomalous divergence of the $U(1)_A$ current and $\mathcal{M}=\text{diag}(\hat m,\hat m, m_s)$ the quark mass matrix.
 In the following, we denote as 
\begin{align}
\pi^a=i\bar\psi_l\gamma_5\tau^a\psi,\quad \delta^a=\bar\psi_l \tau^a \psi_l,\qquad a=1,2,3,
\end{align}
with  $\psi_l$ the light quark doublet, the isotriplet $I=1$ pseudoscalar (pion) and scalar ($a_0(980)$) bilinears,
with 
\begin{align}
\mean{\mathcal{T} \pi^a(x)\pi^b(0)}=&\delta^{ab}P_{\pi\pi} (x), &\mean{\mathcal{T} \delta^a(x)\delta^b(0)}=&\delta^{ab}S_{\delta\delta} (x) \qquad a,b=1,2,3,
\end{align}
 their corresponding euclidean finite-$T$ correlators. Likewise,  
\begin{align}
\eta_l&=i\bar\psi_l \gamma_5 \psi_l,\quad \sigma_l=\bar\psi_l\psi_l,\\ 
\eta_s&=i\bar s \gamma_5 s,\qquad \sigma_s=\bar s s, 
\end{align}
denote the light- and strange-quark part of the isosinglet $I=0$ bilinears, 
with correlators 
\begin{align} 
P_{ll} (x)=&\mean{\mathcal{T} \eta_l(x)\eta_l(0)},&S_{ll} (x)=&\mean{\mathcal{T} \sigma_l(x)\sigma_l(0)},\nonumber\\
P_{ls} (x)=&\mean{\mathcal{T} \eta_l(x)\eta_s(0)},&S_{ls} (x)=&\mean{\mathcal{T} \sigma_l(x)\sigma_s(0)},\label{correlators}\\
P_{ss} (x)=&\mean{\mathcal{T} \eta_s(x)\eta_s(0)},&S_{ss} (x)=&\mean{\mathcal{T} \sigma_s(x)\sigma_s(0)}.\nonumber
\end{align}

 Note that $\eta_l$ and $\eta_s$ mix to give the physical $\eta$ and $\eta'$, while 
the mixing of  $\sigma_l$ and $\sigma_s$ generates the $f_0(500)$ and $f_0(980)$ resonances. These mixings imply that the crossed $ls$ correlators above are nonzero. 

In the same way, the $I=1/2$ pseudoscalar and scalar bilinears are defined as 
\begin{align}
K^a=i\bar\psi\gamma_5\lambda^a\psi,\quad\kappa^a=\bar\psi\lambda^a \psi\qquad a=4,\dots, 7,
\end{align}
respectively, with correlators

\begin{align}
\mean{\mathcal{T} K^a(x)K^b(0)}=&\delta^{ab}P_{KK} (x), &\mean{\mathcal{T} \kappa^a(x)\kappa^b(0)}=&\delta^{ab}S_{\kappa\kappa} (x) \qquad a,b=4,\dots, 7.
\end{align}

From the previous correlators, we will define as usual the associated scalar and pseudoscalar susceptibilities at finite temperature $T$:
\begin{align}
 \chi_Y(T)=\int_T dx  \langle \mathcal{T}\, Y(x) Y (0) \rangle,
\end{align}
with $Y=P,S$ for the different channels discussed above, and $\int_T dx\equiv\intT$. Susceptibilities correspond to the $p=0$ correlators in momentum space and for the particular cases $Y=\bar \psi_l\psi_l$ and $Y=\bar s s$,  subtracting  
$\int_T dx \langle Y\rangle^2$, one gets the mass derivative of the light- and strange-quark condensate, respectively~\cite{GomezNicolaetal}. 

Applying~\eqref{wigen} to a single  bilinear $\mathcal{O}^a=i\bar \psi \gamma_5 \lambda^a\psi\equiv P^a$, 
one obtains  WI relating quark condensates and pseudoscalar susceptibilities~\cite{Nicola:2016jlj}.  
In particular, for our present purposes let us take $\mathcal{O_P}=\pi^b, \eta_l, \eta_s, K^b, A$ in~\eqref{wigen}, which gives:
\begin{align}
&\chi_P^\pi(T)=-\frac{\condl(T)}{\hat m},
\label{wichip}\\ 
&\chi_P^{ ll}(T)=-\frac{\condl(T)}{\hat m}+\frac{m_s}{\sqrt{3}\hat m(\hat m-m_s)}\chi_P^{8A}(T),
\label{wichipetal}\\
&\chi_P^{ss}(T)=-\frac{\conds(T)}{m_s}+\frac{\hat m}{4\sqrt{3}m_s(\hat m - m_s)}\chi_P^{8A}(T),
\label{wichipss}\\
&\chi_P^K(T)=-\frac{\condl (T)+2\conds (T)}{\hat m + m_s},
\label{wichiK}\\
&\chi_P^{AA}=-3\sqrt{3}\frac{\hat m m_s}{m_s-\hat m} \chi_P^{8A},
\label{wiaa}
\end{align}
where $\chi_P^{8A}$ is the susceptibility of the correlator $P_{8A}=\mean{\mathcal{T} P^8(x) A(0)}$. Recall that the basis of $8,0$ states for $I=0$ is related to the $l,s$ state basis as:
\begin{eqnarray}
P^8&=&\frac{1}{\sqrt 3}\left(\eta_l-\eta_s\right), \quad S^8=\frac{1}{\sqrt 3}\left(\sigma_l-\sigma_s\right)\label{PS8},\\
P^0&=&\sqrt{\frac{2}{3}}\left(\eta_l+\eta_s\right), \quad S^0=\sqrt{\frac{2}{3}}\left(\sigma_l+\sigma_s\right)\label{PS0}.
\end{eqnarray}

An additional identity for the correlator $P_{ls}$ in~\eqref{correlators} can be obtained noting from~\eqref{PS8}-\eqref{PS0} that
\begin{align}
P_{ls}=\frac{1}{3}\left(-P_{88}+P_{00}-\frac{1}{\sqrt{2}}P_{08}\right).
\end{align}

Thus, using the WI for $\chi_{88}$, $\chi_{00}$ and $\chi_{08}$ in~\cite{Nicola:2016jlj}, as well as~\eqref{wichip} and~\eqref{wichipetal}, we arrive to
\begin{align}
\chi_P^{ls}(T)=-2\frac{\hat m}{m_s} \chi_{5,disc}(T)=\frac{1}{2\sqrt{3}}\frac{1}{\hat m -m_s}\chi_P^{8A}(T),
\label{wipls}
\end{align}
where $\chi_{5,disc}=\frac{1}{4}\left(\chi_P^\pi-\chi_P^{ll}\right)$ is the parameter customarily used to measure $O(4)\times U(1)_A$ restoration in the lattice~\cite{Buchoff:2013nra}. 
Note that $\pi-\sigma$ and $\sigma-\eta_l$ degeneration for $O(4)$ and $U(1)_A$ restoration, respectively, implies the vanishing of $\chi_{5,disc}$ at the $O(4)\times U(1)_A$ transition (see section \ref{sec:cons} for details).
In addition, $\chi_{5,disc}$ is proportional to the topological susceptibility, as we will discuss in detail in section~\ref{sec:I01}.   
The relations~\eqref{wipls} would be  testable in lattice and as we will see in section~\ref{sec:latt}, play an important role regarding chiral pattern restoration.

Let us now consider the generic identity~\eqref{wigenvec}, which  becomes non trivial only in the $I=1/2$ channel. Thus, taking $\mathcal{O}^b=\kappa^b$, the only surviving term in the l.h.s of~\eqref{wigenvec} corresponds to 
combinations $\delta{\cal O}^b(y)/\delta\alpha_V^a(x)=\delta(x-y)f_{ab8}S^8$ with $a,b=4,\dots, 7$ and $f_{abc}$ the $SU(3)$ antisymmetric structure constants. The l.h.s of~\eqref{wigenvec}  becomes then a combination of the light and strange quark condensates, while the r.h.s. gives rise to the scalar susceptibility $\chi_S^\kappa$.  Since $f_{458}=f_{678}=\sqrt{3}/2$, we finally obtain: 

\begin{equation}
 \chi_S^\kappa (T)=\frac{\condl (T)-2\conds (T)}{m_s-\hat m}.
 \label{wichikappa}
 \end{equation}

Combining this new identity with the kaon WI in~\eqref{wichiK} gives
\begin{equation}
  \chi_S^\kappa (T)-\chi_P^K (T)=\frac{2}{m_s^2-\hat m^2}\left[m_s\condl (T)-2\hat m \conds (T)\right],
  \label{wichikappakaondif}
\end{equation}
which establishes a relation for $K$-$\kappa$ degeneracy to be analyzed  in section~\ref{sec:I1/2}.  

We can obtain new additional WI by considering in~\eqref{wigen} a two-point function $\mathcal{O}^{ab}=P^aS^b$, with $P^a$ and $S^b$ generic pseudoscalar and scalar bilinears connected by $SU(2)_A$ transformations, e.g. $P^a=\pi^a, S^b=\sigma$ and so on. Expressing the results in the basis of $l,s$ correlators through~\eqref{PS8}-\eqref{PS0} we get
\begin{align}
P_{\pi\pi}(y)-S_{ll}(y)=\hat{m} \int_T dx \mean{\mathcal{T} \sigma_l(y)\pi(x)\pi(0)},\label{wipisigmaop}\\
P_{ll}(y)-S_{\delta\delta}(y)=\hat{m} \int_T dx \mean{\mathcal{T} \delta(y)\pi(x)\eta_l(0)},\label{wideltaetaop}\\
P_{ls}(y)=\frac{1}{3}\hat{m} \int_T dx \mean{\mathcal{T} \eta_s(y)\pi(x)\delta(0)},\label{wiPls}\\
S_{ls} (y)=-\frac{1}{3}\hat{m} \int_T dx \mean{\mathcal{T} \sigma_s(y)\pi(x)\pi(0)},\label{wiSls}\\
d^{abc}\left[P_{KK}(y)-S_{\kappa\kappa}(y)\right] =\hat{m} \int_T dx \mean{\mathcal{T} K^b(y)\kappa^c(x)\pi^a(0)},\label{wiKKappaop}
\end{align}
with  $d^{abc}$ the symmetric $SU(3)$ coefficients, $a=1,2,3$ and $b,c=4,\dots,7$. These equations parameterize the degeneracy of $SU(2)_A$ chiral partners in terms of three-point functions; 
the latter encode the physical vertices responsible for the breaking of such a degeneracy. 
Furthermore, if $P^a$ and $S^b$ are bilinears linked now through a $U(1)_A$ transformation, equation ~\eqref{wigen} gives rise to  
\begin{eqnarray}
P_{\pi\pi}(y)-S_{\delta\delta}(y)&=&\int_T dx \mean{\mathcal{T}\pi(y)\delta(0)\tilde\eta(x)},\label{wipideltaop}\\
P_{ll}(y)-S_{ll}(y)&=&\int_T dx \mean{\mathcal{T}\eta_l(y)\sigma_l(0)\tilde\eta(x)},\label{wietasigmaop}\\
P_{ls}(y)-S_{ls}(y)&=&\int_T dx \mean{\mathcal{T}\eta_l(y)\sigma_s(0)\tilde\eta(x)},\label{PlsminusSlsWI}\\
P_{ss}(y)-S_{ss}(y)&=&\int_T dx \mean{\mathcal{T}\eta_s(y)\sigma_s(0)\tilde\eta(x)},\label{PssminusSssWI}\\
P_{KK}(y)-S_{\kappa\kappa}(y)&=&\int_T dx \mean{\mathcal{T}K(y)\kappa(0)\tilde\eta (x)},\label{wiKKappaopS}
\end{eqnarray}
where $\tilde\eta(x)=\hat{m}\eta_l(x)+m_s\eta_s(x)+\frac{1}{2}A(x).$
The above equations include now explicit $m_s$ and anomalous ($A$) terms responsible for $U(1)_A$ breaking.  The possible implications of identities~\eqref{wipisigmaop}-\eqref{wiKKappaopS} regarding $O(4)$ and $U(1)_A$ restoration, as well as their connection with meson scattering processes, are discussed in section~\ref{sec:2p3p}. 

All the identities in this section have been  formally derived from the QCD generating functional. Hence, up to renormalization ambiguities related to the fields and vertices involved~\cite{Bochicchio:1985xa, Boucaud:2009kv}, 
they should be respected by any model or lattice calculation. In fact, the one-point WI relating quark condensates and pseudoscalar susceptibilities have been verified recently 
in the hadronic sector through Chiral Perturbation Theory (ChPT)~\cite{Nicola:2016jlj} and the identities~\eqref{wichip} and~\eqref{wichipss} have been tested in the lattice~\cite{Buchoff:2013nra}.

In addition, these WI should be valid irregardless of the regime  of symmetry restoration. 
In the next section, we will exploit that feature by considering  symmetry transformations of the different correlators and show that this leads to rather strong conditions on $O(4)$ and $U(1)_A$ partner degeneration. 
In this way, the results derived in this work only make use of symmetry arguments and hence are valid independently of the representation used.

\section{Consequences for chiral symmetry restoration}
\label{sec:cons}

In this section, we will analyze the consequences of the WI derived in section~\ref{sec:wi} regarding the behavior of chiral patterns and partners. First, let us briefly review how the different bilinears in the $I=0,1$ sectors and their correlators are connected through infinitesimal $O(4)$ and $U(1)_A$ transformations. Similar transformations for the $I=1/2$ sector will be discussed below. 

On the one hand, $SU(2)_A$ transformations mix $\pi-\sigma_l$ and $\delta-\eta_l$ states, namely, 

\begin{align}
&\delta \pi^a (y)/\delta\alpha_A^b(x)=-\delta_{ab}\delta (x-y)\sigma_l(x),&&\delta\sigma_l(y)/\delta\alpha_A^b(x)=\delta(x-y)\pi^b(x)\nonumber\\  
&\delta \delta^a (y)/\delta\alpha_A^b(x)=\delta_{ab}\delta (x-y)\eta_l(x),&&\delta\eta_l(y)/\delta\alpha_A^b(x)=-\delta(x-y)\delta^b(x). 
\label{chitransbi}
\end{align}
with $a,b=1,2,3$. 

Thus, if chiral symmetry is  restored, one can rotate $\pi$ to $\sigma_l$ and $\eta_l$ to $\delta$, so that their correlators become degenerate in a $O(4)$ restoration scenario. 
A specific transformation for such rotation is discussed below in section~\ref{sec:I01}. 
On the other hand,  $U(1)_A$ rotations allow one to connect bilinears with the same isospin but opposite parity, namely $\pi-\delta$ and $\eta_l-\sigma_l$, 

\begin{align} 
&\delta\pi^{a}(y)/\delta\alpha_A(x)=-\delta(x-y)\delta^a(x),&&\delta\delta^{a}(y)/\delta\alpha_A(x)=\delta(x-y)\pi^a (x),\nonumber\\ 
&\delta\sigma_l(y)/\delta\alpha_A(x)=\delta(x-y)\eta_l (x),&&\delta\eta_l(y)/\delta\alpha_A(x)=-\delta(x-y)\sigma_l (x),
\label{diagtrans}
\end{align}
with $\alpha_A=\sqrt{2/3}\alpha_A^0$, 
which would become degenerate $U(1)_A$ partners. As explained above, we define $U(1)_A$ restoration as the regime where those partners are approximately degenerated. 
In summary, chiral partners in this sector are related through
\begin{align}
P_{\pi\pi}\eqchiral S_{ll}, \quad P_{ll}\eqchiral S_{\delta\delta},\label{partnerso4}\\
P_{\pi\pi}\equa S_{\delta\delta}, \quad P_{ll}\equa S_{ll},
\label{partnersua1}
\end{align}
and so on for their corresponding susceptibilities. 

The four partners in~\eqref{partnerso4}-\eqref{partnersua1} would become degenerate within a $O(4)\times U(1)_A$ pattern where $\chi_{5,disc}\sim 0$.  
In turn, in a full $U(3)$ restoring scenario all members of the scalar/pseudoscalar nonets would become degenerate. 
Nevertheless, the latter limit is meant to be reached at a much higher temperature than chiral restoration, as seen for instance in the degeneration of screening masses~\cite{Cheng:2010fe}. 
The four correlators~\eqref{partnerso4}-\eqref{partnersua1} have been actively investigated in  lattice and theoretical analyses 
to study partner degeneracy~\cite{Buchoff:2013nra,Ishii:2015ira,Ishii:2016dln,Cossu:2013uua,Tomiya:2016jwr,Brandt:2016daq,Nicola:2013vma,GomezNicola:2016csg}.  

Next, we will explain in detail the main results of our present work, which arise as consequences of the previous WI analysis.

\subsection{$I=0,1$ sectors: $O(4)$ vs $U(1)_A$ restoration}
\label{sec:I01}

First of all, we will show how the analysis of the crossed $ls$ correlators allows one to reach conclusions regarding chiral and $U(1)_A$ restoration.
Note that they are nonzero below the phase transition due to the $\eta-\eta^\prime$ and $f_0(500)-f_0(980)$ mixing.
In fact, the mixing in the pseudoscalar sector is still present at the light chiral limit~\cite{Guoetal} and $\chi_P^{8A}$ is nonzero at $T=0$ when $\hat m=0$~\cite{Nicola:2016jlj} (see our comments below). 

However, the $ls$ correlators should be exactly zero at $O(4)$ restoration, i.e. in the regime where the $O(4)$ chiral partners in~\eqref{partnerso4} degenerate.  
The reason is as follows: a general $SU(2)_A$ transformation acting on the $\eta_l$ bilinear
\begin{equation}
\eta_l(x)\rightarrow i\bar\psi_l(x)\gamma_5 e^{i\gamma_5 \alpha_a\tau^a}\psi_l(x)=i\bar\psi_l(x)\gamma_5 \cos (\alpha_a\tau^a)\psi_l(x)-\bar\psi_l(x) \sin (\alpha_a\tau^a)\psi_l(x),
\label{chilsvanishing1}
\end{equation}
with $a=1,2,3$ can be written as a sum of $P$-odd (first) and $P$-even (second) contributions. 
Thus, it is possible to choose an angle $\alpha_a$  in~\eqref{chilsvanishing1} such that the $P$-odd part vanishes. 
For instance, chosing for any $b=1,2,3$
\begin{equation}
\alpha_b=\pi/2 \quad{\rm and}\quad \alpha_{a\neq b}=0 \qquad 
\label{trans}
\end{equation}
we get
\begin{equation}
\eta_l(x)\rightarrow -\bar\psi_l(x) \tau^b \psi_l(x)=-\delta^b (x)\Rightarrow P_{ls}(x)\rightarrow -\mean {{\cal T}  \delta^b (x) \eta_s(0)} =0,
\label{chilsvanishing2}
\end{equation}
where we have used that $\eta_s$  is invariant under $SU(2)_A$ transformations and the last correlator vanishes by parity. 
Thus,  since expectation values of transformed fields should equal the untransformed ones if the symmetry is exact, in the regime where $SU(2)_A$ restoration is effective  $P_{ls}\rightarrow 0$,
Similarly, we obtain $S_{ls}\rightarrow 0$ as the system approaches the chiral transition.  Therefore, following our notation, we have showed that
\begin{equation}
P_{ls}\eqchiral 0, \qquad S_{ls}\eqchiral 0.
\label{lscond}
\end{equation}

Note that this is actually the same argument that leads to $\condl=0$ for exact chiral restoration ($\condl\eqchiral 0$ in our notation) since the  transformations~\eqref{trans} rotate $\condl=\mean{\sigma_l}\rightarrow \mean{\pi^b}=0$ by parity. 
In this case, $\condl\eqchiral 0$ relies on $\sigma_l-\pi$ degeneration at chiral restoration, while~\eqref{lscond} is a consequence of the $\delta-\eta_l$ one. 

In the same way, considering a pure $U(1)_A$ transformation acting on the bilinears
\begin{eqnarray}
\eta_l(x)&\rightarrow& i\bar\psi_l(x)\gamma_5 e^{i\gamma_5 \alpha_A}\psi_l(x)=\cos (\alpha_A) \eta_l-\sin(\alpha_A) \sigma_l,\nonumber\\
\eta_s(x)&\rightarrow& i\bar s (x) \gamma_5 e^{i\gamma_5 \alpha_A} s(x)=\cos(\alpha_A) \eta_s - \sin(\alpha_A) \sigma_s
\end{eqnarray}
and choosing as before $\alpha_A=\pi/2$, we have $P_{ls}\rightarrow S_{ls}$ and $P_{ss}\rightarrow S_{ss}$. 
Thus, at $U(1)_A$ restoration we obtain
\begin{align}
P_{ls}\equa S_{ls}, \quad P_{ss}\equa S_{ss}, 
\label{ua1deg}
\end{align}

The additional chiral restoring conditions in~\eqref{lscond}~and~\eqref{ua1deg} could be indeed tested in the lattice (see section~\ref{sec:latt}). 
Nevertheless, the main consequences of these results for the  pattern of chiral restoration is highlighted when~\eqref{lscond}  is used in connection with the WI~\eqref{wipls}.
Since~\eqref{lscond} implies $\chi_P^{ls}\eqchiral 0$ at chiral restoration, or more precisely for exact degeneration of $\delta-\eta_l$~\eqref{chilsvanishing1}-\eqref{chilsvanishing2}, 
the combination of~\eqref{wipls} and~\eqref{lscond} leads to the following conclusion:
\begin{equation}
\chi_P^{ll}\eqchiral\chi_S^{\delta} \ \Rightarrow \chi_P^{ls}\eqchiral 0 \ \Rightarrow  \chi_{5,disc} \eqchiral 0.
\label{chi5degchiral}
\end{equation}

Therefore, our WI analysis supports $U(1)_A$ partner degeneration if $O(4)$ partners {\em exactly} degenerate, which is a central result of this work.  More specifically, as mentioned above,  $\chi_{5,disc}$ is a suitable parameter to measure $O(4)\times U(1)_A$ restoration in terms of the $\pi-\eta_l$ partner degeneration, while the l.h.s. of~\eqref{chi5degchiral} relies on the chiral $O(4)$ degeneration of $\delta-\eta_l$ partners. 
This point will be relevant for the analysis of lattice results in the crossover regime, where not all $O(4)$ partners need to degenerate at the same temperature. 

Another argument that provides further support to our previous conclusion is connected with the topological susceptibility, 
 defined as the correlator of the anomaly operator 
\footnote{The normalization factor~\eqref{chitopdef} is chosen so that the definition of $\chi_{top}$ coincides with~\cite{Buchoff:2013nra}. Such factor comes  from our normalization of $A(x)$ and our definition of Euclidean gauge fields, which follows~\cite{Nicola:2016jlj}.} 
\begin{equation}
\chi_{top}(T)\equiv -\frac{1}{36}\chi_P^{AA}(T)=-\frac{1}{36}\int_T dx  \langle \mathcal{T} A(x) A(0) \rangle. 
\label{chitopdef}
\end{equation} 
Since $\chi_{top}$  is the correlator of the topological density, whose charge measures the difference between left-handed and right-handed zero modes of the Dirac operator (Atiyah-Singer index theorem), 
it provides a direct measure of  $U(1)_A$ breaking.
Although $\chi_{top}$ is particularly difficult to measure in the lattice~\cite{Buchoff:2013nra,Borsanyi:2016ksw,Petreczky:2016vrs}, its vanishing or asymptotic reduction indicate  $U(1)_A$ restoration, since the system becomes less sensitive to the $P$-breaking anomaly contribution parametrized in the $\theta$-term~\cite{ Azcoiti:2016zbi}. 

Here,  we will make use once more of the WI derived in section~\ref{sec:wi} to reach specific conclusions about $\chi_{top}$. Thus, combining~\eqref{wiaa} and~\eqref{wipls} we obtain
\begin{equation}
\chi_{top}(T)=\hat m ^2 \chi_{5,disc}(T),
\label{chi5vschitop}
\end{equation}
which was also derived in~\cite{Buchoff:2013nra} from the properties of the Dirac operator and is therefore a consistency check for the WI derived here. 
In addition, from~\eqref{chi5vschitop} and~\eqref{wipls}  we  can conclude that
\begin{equation}
\chi_P^{ls}(T)=-\frac{2}{{\hat m}\,m_s} \chi_{top}(T).
\label{chilsvschitop}
\end{equation}
Thus, using~\eqref{chi5degchiral},
\begin{equation}
\chi_P^{ll}\eqchiral\chi_S^{\delta} \ \Rightarrow \ \chi_{top} \eqchiral 0,
\label{chitopvan}
\end{equation}
i.e, the topological susceptibility should also vanish at the temperature regime where  $O(4)$ partners exactly degenerate. 
The same conclusion about the vanishing of $\chi_{top}$ for any temperature above chiral restoration has been reached in~\cite{ Azcoiti:2016zbi}. Note that the main argument in that work actually relies on the identity 
\begin{equation}
\chi_P^{ ll}(T)=-\frac{\condl(T)}{\hat m}-\frac{4}{\hat m^2}\chi_{top}(T),
\end{equation}
which is nothing but the combination of our identities~\eqref{wichipetal} and \eqref{wiaa}, using~\eqref{chitopdef}. Therefore, our results here are fully consistent with~\cite{ Azcoiti:2016zbi}. 

Let us remark that in the light chiral limit $\hat m\rightarrow 0^+$, neither $\hat m\,\chi_{5,disc}$ nor $\chi_P^{8A}$ in~\eqref{wipls} vanish at $T=0$. 
In fact, the latter vanishes at $T=0$ only when the anomalous part of the $\eta'$ mass goes to zero for fixed $\hat m$~\cite{Nicola:2016jlj}~{\footnote{There is a missing multiplying $M_0^2$ in the LO ChPT expression for $\chi_P^{8A}$ in eq.(A.3) in~\cite{Nicola:2016jlj}.}. Therefore, we expect $\chi_{5,disc}\sim 1/{\hat m}$ and $\chi_{top}\sim {\hat m}$  away from $T_c$ (the latter from~\eqref{chi5vschitop}). This is supported also by~\cite{ Azcoiti:2016zbi}, where it is argued that $\chi_{top}\sim {\hat m}\condl$ in the chiral limit, since $\condl$ is regular in that limit~\cite{Smilga:1995qf,Ejiri:2009ac}. Hence, the vanishing  of $\chi_{5,disc}$ in~\eqref{chi5degchiral} and $\chi_{top}(T)$ in~\eqref{chitopvan} are genuine consequences of chiral restoration, which ideally would require $\hat m\rightarrow 0^+$ {\it and} $T\rightarrow T_c$.  
The key point is that for any value of the light quark mass $\hat m$,~\eqref{wipls} and~\eqref{chilsvschitop} connect the chiral $O(4)$ restoring observable $\chi_P^{ls}$, with $U(1)_A$-restoring ones, $\chi_{5,disc}$ and $\chi_{top}$.

\subsection{$I=0,1$ sectors: connection with lattice results}
\label{sec:latt}

Let us now comment on  the connection of our previous analysis with lattice results.  
On the one hand, our main result in section~\ref{sec:I01}, i.e. $U(1)_A$ partner degeneration as a consequence of chiral restoration, is consistent with the $N_f=2$ lattice results near the chiral limit in~\cite{Cossu:2013uua,Tomiya:2016jwr} and for physical pion masses in~\cite{Brandt:2016daq}. In particular, $\delta-\eta_l$ degeneration is very effective at chiral restoration in the latter work, which according to~\eqref{chi5degchiral} explains the $U(1)_A$ $\pi-\eta_l$ degeneration that they find near chiral restoration.  
On the other hand, the $N_f=2+1$ lattice analysis in~\cite{Buchoff:2013nra} supports  $U(1)_A$ partners to degenerate at a higher temperature than $O(4)$ ones. 

Let us remark once more that our conclusions in~\eqref{chi5degchiral} and~\eqref{chitopvan} stand on O(4) degeneration, and hence they should be more accurate near the light chiral limit and for two flavors. 
Physical masses and strange-quark effects may distort numerically this picture. 
In addition, the numerical values for $O(4)$ partner degeneration in~\cite{Buchoff:2013nra} (Table IV in that paper) show that the thermal evolution of the difference $\chi_P^{ll}-\chi_S^\delta$, although with large errors, does not reduce significantly around $T_c$. In fact, that difference remains sizable up to the region where the $U(1)_A$ is approximately restored, i.e. where $\chi_P^{\pi}$ and $\chi_S^{\delta}$ degenerate.
Recall that $\eta_l-\delta$ degeneration is indeed the main assumption in our previous argument. 
 Thus, the absence of strange quark corrections in the $N_f=2$ lattice analyses in~\cite{Cossu:2013uua,Tomiya:2016jwr,Brandt:2016daq} may explain why they obtain a $O(4)\times U(1)_A$ pattern, consistently with our conclusions~\eqref{chi5degchiral} and ~\eqref{chitopvan}.

From the previous considerations, it would be natural to expect in the real world a relation between $\chi_{5,disc}$ and typical chiral-restoring order parameters.  
Obviously, the most natural candidate is the light-quark condensate, and, consequently, one could assume that the temperature scaling of $\chi_{5,disc}$ is dictated by some positive power of $\condl$, up to corrections in the light quark mass.  As mentioned before, that scaling is also consistent with the behaviour $\chi_{top}\sim {\hat m}\condl$ found in~\cite{ Azcoiti:2016zbi} in the chiral limit, and the relation~\eqref{chi5vschitop}.  To test this assumption, we compare in Fig.~\ref{fig:latticenonet}a the $T$ scaling of  the lattice data in~\cite{Buchoff:2013nra} for $\chi_{5,disc}$ with   the subtracted quark condensate

\begin{equation}
\Delta_{l,s}(T)=\condl (T)-2(\hat m/m_s)\conds (T),  
\label{subcond}
\end{equation}
which is customarily used as order parameter in lattice analyses to avoid finite-size divergences  $\mean{\bar q_i q_i}\sim m_i/a$~\cite{Buchoff:2013nra}.   This plot shows that the $\chi_{5,disc}$ scaling fits reasonably well between those for $\Delta_{l,s}$ and $\Delta_{l,s}^{1/2}$. The latter is motivated  by  considering a simple realization of the quark bilinear $\pi^a$ in terms of a pion field $\tilde\pi^a$ in a meson lagrangian, as far as their  expectation values are concerned, through a normalization constant $\pi^a=N_\pi\tilde\pi^a$. Then, $N_\pi^2=-\condl G_\pi^{-1}(0)/\hat m$ from the WI~\eqref{wichip} with $G_\pi(p)$ the pion propagator. Therefore, we would get such $\sqrt{\condl}$ scaling from~\eqref{wiPls} assuming a smooth dependence of the pion self-energy,  which does not show any critical behavior. 

\begin{figure*}[h]
\centerline{
\includegraphics[width=8cm]{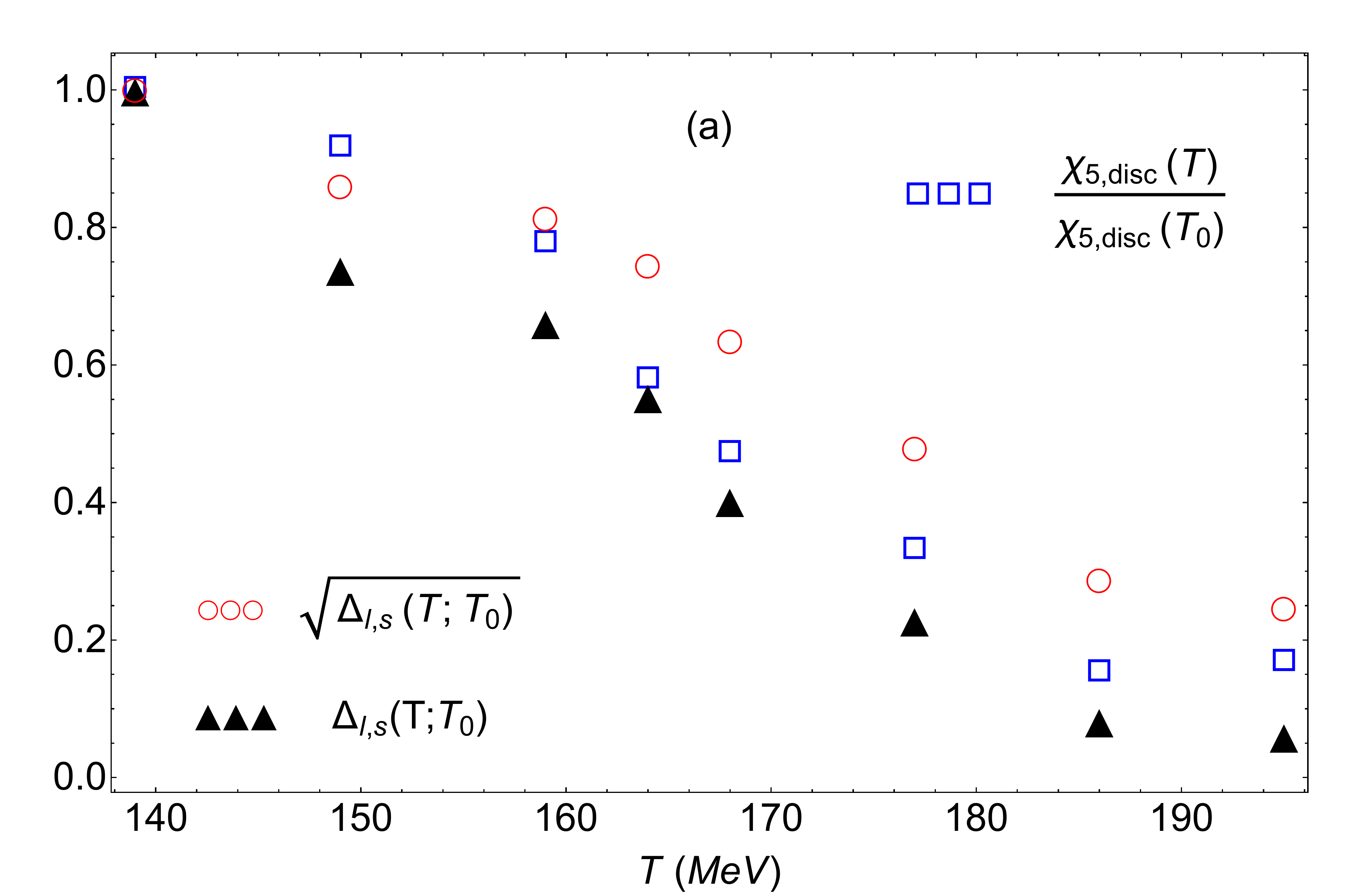}
\includegraphics[width=8cm]{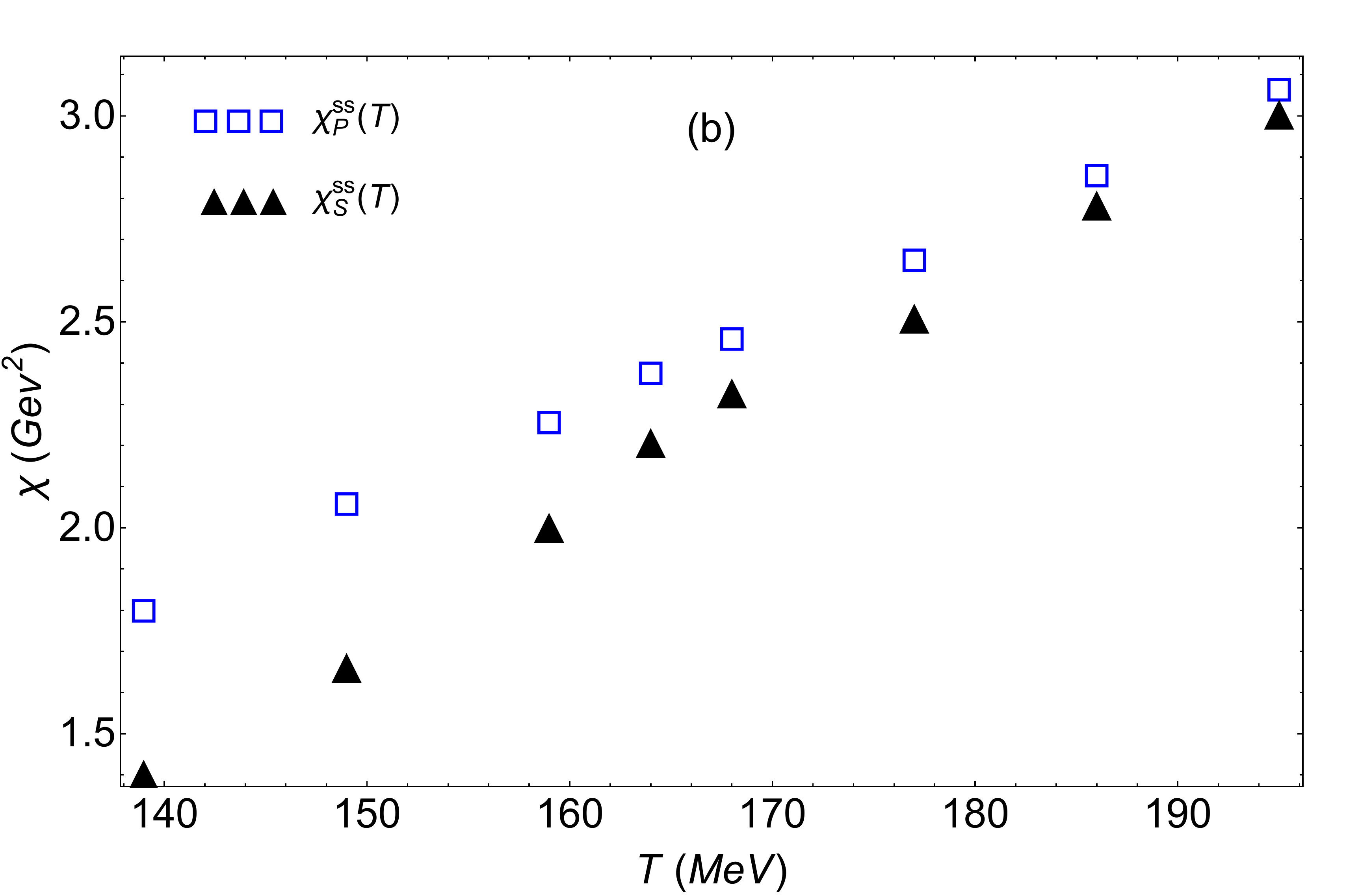}}
\centerline{
\includegraphics[width=8cm]{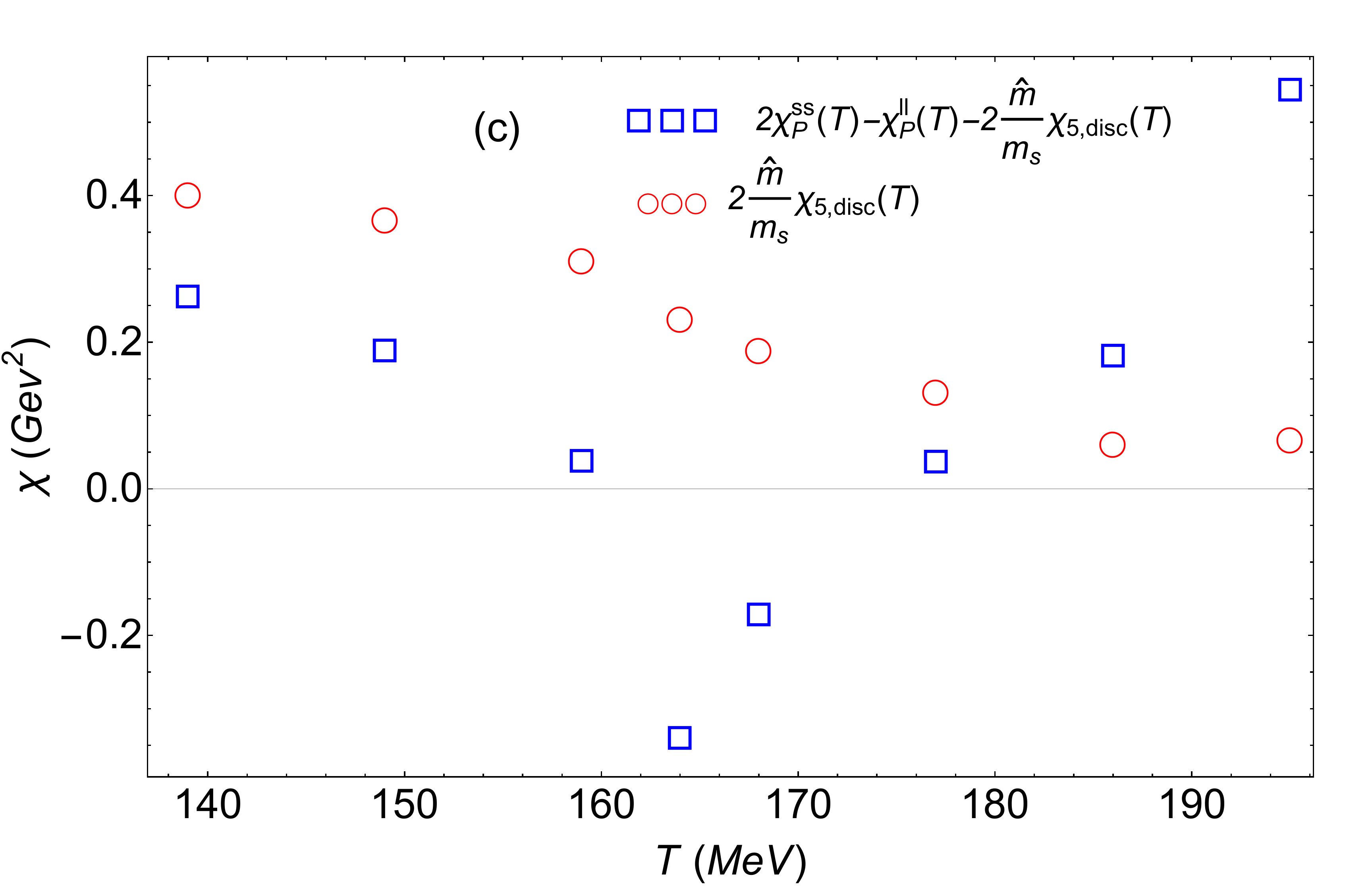}
\includegraphics[width=8cm]{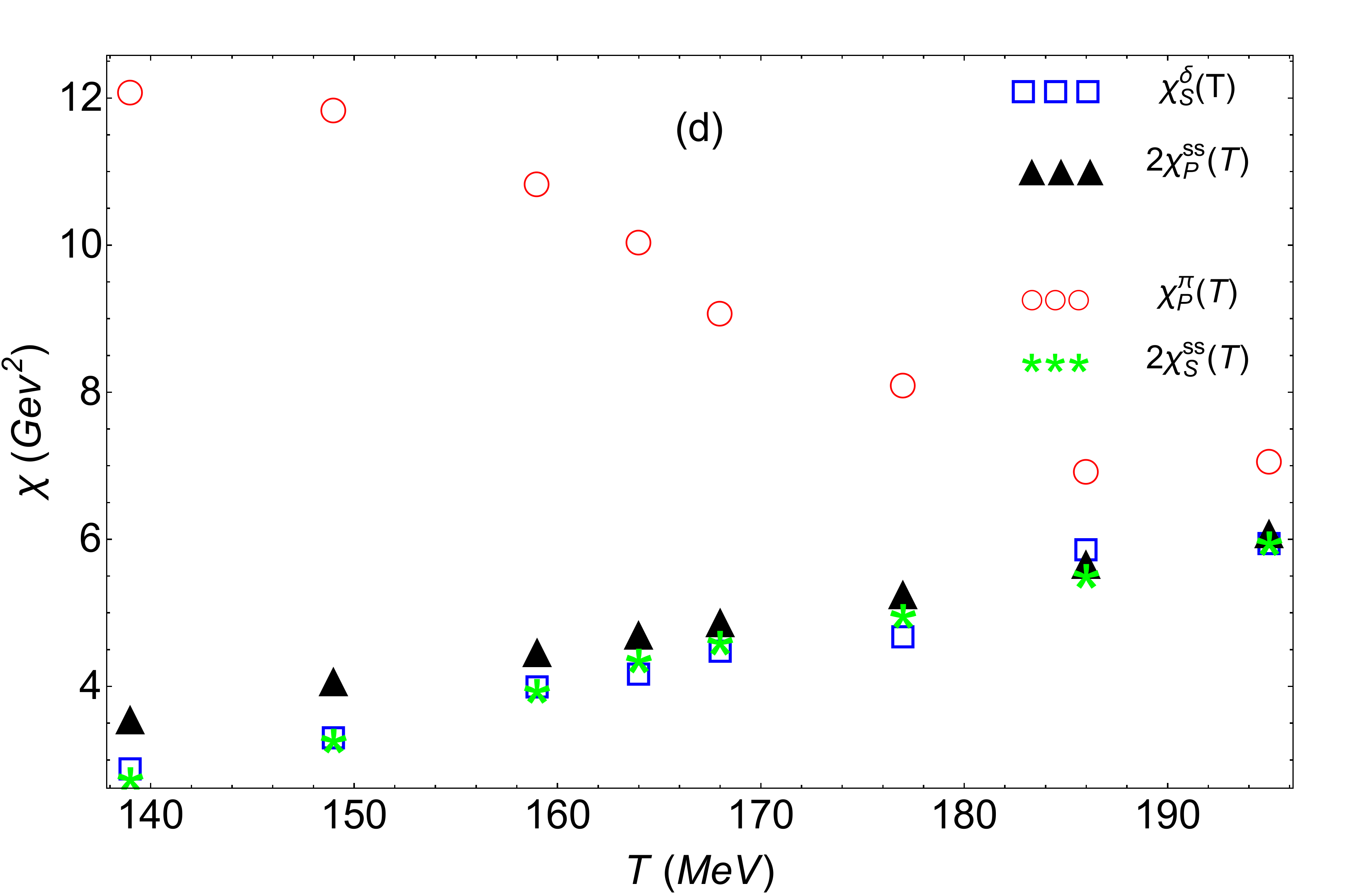}
}
\caption{Different susceptibilities combinations from the lattice data in~\cite{Buchoff:2013nra} for $32^3\times 8$ lattice size. (a): Comparison between the scaling of $\chi_{5,disc}$ and the subtracted quark condensate $\Delta_{l,s}(T;T_0)=\Delta_{l,s}(T)/\Delta_{l,s}(T_0)$ with respect to the reference temperature $T_0=139$ MeV. (b): Scalar and Pseudo-scalar pure strange susceptibilities. (c): Susceptibility combination related to the vanishing of the $\eta-\eta'$ mixing angle with $\hat m/m_s=0.088$~\cite{Buchoff:2013nra}, where we also plot $-\chi_P^{ls}$ according to~\eqref{wipls}. (d): Partner degeneracy in the  scenario where the two parameters in (c) remain small.}
\label{fig:latticenonet}
\end{figure*}

In addition, the vanishing of $\chi_{5,disc}$ at $O(4)\times U(1)_A$ restoration also implies the vanishing of $\chi_{S}^{disc}$, 
i.e. the disconnected part of the scalar susceptibility $\chi_S^{ll}$ since $\chi_{5,disc}\sim\chi_{S}^{disc}$ in this limit~\cite{Buchoff:2013nra}. 
Note that this is not in contradiction with the expectation of a scalar susceptibility peak at the critical point in the light chiral limit, which only applies to the total susceptibility~\cite{Smilga:1995qf,Ejiri:2009ac}.
Consequently, its connected part $\chi_{S}^{con}$ would also peak at  $O(4)\times U(1)_A$ restoration.  
Actually, in the physical massive case, $\chi_{S}^{con}=\chi_S^\delta/2$ grows with $T$ below the transition~\cite{Buchoff:2013nra,Nicola:2011gq,GomezNicola:2016csg} and, since $\chi_P^\pi$ decreases like $\condl$~\cite{Nicola:2016jlj}, 
their degeneracy would give rise to a maximum for $\chi_S^\delta$ near $U_A(1)$ restoration. 
A hint of that behavior for $\chi_{S}^{con}$ is seen at around $T\sim$ 190 MeV in the lattice~\cite{Bazavov:2011nk}. 
This is indeed consistent with the region of $U(1)_A$ restoration obtained by that collaboration, albeit in their more recent analysis~\cite{Buchoff:2013nra} higher $T$ data points would be needed.

Finally the degeneracy conditions in~\eqref{ua1deg} can be checked in the lattice using the data in~\cite{Buchoff:2013nra} for the $\bar ss$ channel. 
The comparison is depicted in Fig.~\ref{fig:latticenonet}b and shows a clear sign of degeneracy around the asymptotic $U(1)_A$ restoration regime in~\cite{Buchoff:2013nra}, confirming our present analysis.

\subsection{$I=1/2$ sector: $K$ vs $\kappa$ degeneration}
\label{sec:I1/2}

Our analysis leads also to interesting consequences for the $I=1/2$ sector. 
From the WI in~\eqref{wichikappakaondif}, $K$-$\kappa$ degeneration at  chiral restoration requires the light chiral condensate and the light quark mass both to vanish
\begin{equation}
\condl\eqchiral 0, \hat m\to 0 \Rightarrow \chi_S^\kappa (T)\eqchiral\chi_P^K (T).
\label{equivkaonkappa}
\end{equation}

This chiral partner degeneration is also consistent with the choice of a $SU(2)_A$ rotation for the $K$ and $\kappa$ correlators. Infinitesimally, one has 
\begin{align}
\delta K^a (y)/\delta \alpha_A^b (x)&=-\delta(x-y)d_{abc}\kappa^c(x),& \delta \kappa^a (y)/\delta \alpha_A^b (x)&=\delta(x-y)d_{abc}K^c(x) ,
\label{kaoninf}
\end{align}
with $a,c=4,\dots,7$ and $b=1,2,3$.

Let us  show now that one can indeed choose a chiral rotation  that transforms $P_{KK}$ into a pure $S_{\kappa\kappa}$ correlator.  Taking one of the three $SU(2)_A$ angles $\alpha_b\neq 0$ with $\alpha_{a\neq b}=0$ ($a,b=1,2,3$) allows one to write

\begin{equation}
e^{i\gamma_5\alpha_a\lambda^a/2}=\ID-\ID_2+\cos(\alpha_b/2)\ID_2+i\sin(\alpha_b/2)\gamma_5\lambda_b,
\end{equation}
where $\ID_2=\diag(1,1,0)=\frac{\sqrt{3}}{3}(\sqrt{2}\lambda^0+\lambda^8)$. Taking into account  that $\ID_2\lambda^c\ID_2=\lambda^b\lambda^c\lambda^b=\ID_2\lambda^b\lambda^c=\lambda^b\lambda^c\ID_2=0$, $\{\ID_2,\lambda^b\}=\lambda^b$, $\{\lambda^b,\lambda^c\}=2d^{bce}\lambda^e$ with   $b=1,2,3$, $c=4,\dots,7$, under such rotation, we obtain

\begin{equation}
K^c\rightarrow \cos(\alpha_b/2) K^b - 2d_{bce}\sin(\alpha_b/2)\kappa^e,
\label{partktrans}
\end{equation}
which for infinitesimal local transformations reduces to the first equation of~\eqref{kaoninf}.  Thus, taking into account that  the nonvanishing $d_{bce}=\pm 1/2$ for $b=1,2,3$, $c,e=4,\dots,7$, we conclude that setting $\alpha_b=\pi$ in~\eqref{partktrans} yields

\begin{equation}
P_{KK}\eqchiral S_{\kappa\kappa}
\label{Kkappachiral}
\end{equation}
consistently with~\eqref{wichikappakaondif} and~\eqref{equivkaonkappa}.

The result~\eqref{equivkaonkappa}, as it happened with~\eqref{chi5degchiral} and~\eqref{chitopvan}, is valid only in the exact chiral restoration regime. Nevertheless, we can take further advantage of the WI~\eqref{wichikappakaondif} also in the physical crossover regime, by writing that identity in terms of the subtracted condensate defined in~\eqref{subcond}
\begin{equation}
  \chi_S^\kappa (T)-\chi_P^K (T)=\frac{2m_s}{m_s^2-\hat m^2}\Delta_{l,s}(T).
  \label{wichikappakaondif2}
\end{equation}

Therefore, our analysis not only establishes the degeneracy of $K-\kappa$ partners in this sector but provides a direct way to measure the breaking of that degeneracy in the lattice through~\eqref{wichikappakaondif2}.  
This is another important result of the present work.  Recall that numerically, the value of $\Delta_{ls}$ is reduced by one half at the chiral transition  with respect to the $T=0$ value~\cite{Bazavov:2011nk,Buchoff:2013nra}.   
The asymptotic $K$-$\kappa$ degeneracy observed for lattice screening masses~\cite{Cheng:2010fe} is also consistent with this conclusion. It must  also be kept in mind that $K-\kappa$ correlators can  be connected  as well  by $U(1)_A$ rotations, which infinitesimally read

\begin{align}
\delta K^a (y)/\delta \alpha_A^0 (x)&=-\sqrt{2/3}\delta(x-y)\kappa^a(x),& \delta \kappa^a (y)/\delta \alpha_A^0(x)&=\sqrt{2/3}\delta(x-y)K^a(x),
\end{align}
with $a=4,\dots,7$. Under a general $U(1)_A$ transformation,

\begin{equation}
K^c\rightarrow \cos(\alpha_A)K^c-\sin(\alpha_A)\kappa^c
\end{equation}
so that choosing $\alpha_A=\pi/2$, we conclude:

\begin{equation}
P_{KK}\equa S_{\kappa\kappa}.
\label{Kkappaua1}
\end{equation}

Thus, in an ideal chiral-restoring scenario,~\eqref{Kkappachiral} and~\eqref{Kkappaua1} are consistent with our results in section~\ref{sec:I01} where $O(4)$ and $U(1)_A$ restoration coexist, while, in the physical crossover case, the $U(1)_A$ restoring effects beyond chiral restoration will also contribute to $K-\kappa$ degeneration. In any case the degeneration would be  parametrized by $\Delta_{l,s}$ through~\eqref{wichikappakaondif2}.

\subsection{Mixing angles}

Our  WI analysis also provides relevant conclusions regarding $\eta-\eta'$ mixing.  For that purpose, we define the mixing angle  in the standard fashion, which will be enough for our present discussion. 
The mass eigenstates $\eta$ and $\eta^\prime$ are defined from the flavor eigenstates $\eta_8$ and $\eta_0$ by 

\begin{eqnarray}
\eta=\eta_8\cos\theta_P -\eta_0\sin\theta_P,  \nonumber\\
\eta'=\eta_8\sin\theta_P  +\eta_0\cos\theta_P, 
\label{mix}
\end{eqnarray}
and so on in the scalar sector with the replacements $\theta_P\rightarrow \theta_S$, $\eta\rightarrow f_0(500)$ and $\eta'\rightarrow f_0(980)$. 
The  mixing angles and correlators are meant to be temperature dependent.  By definition, the mixing angle $\theta_P(T)$ is defined to cancel the correlator
\begin{align}
P_{\eta\eta'}=\frac{1}{6}\left(2P_{ss}-P_{ll}-8P_{ls}\right)\sin2\theta_P  
+\frac{\sqrt{2}}{3}\left(P_{ll}-2P_{ss}-P_{ls}\right) \cos 2\theta_P=0.
\label{mixcond}
\end{align}
where we have used the relation between the $P_{0,8}$ and $P_{l,s}$ correlators,  as we did in section~\ref{sec:wi}. 

A  relevant limit  is ideal mixing,  $\sin\theta_P^{id}=-\sqrt{2/3}$, so that $\eta\sim \eta_l$, $\eta'\sim\sqrt{2}\eta_s$. 
That limit is reached at $T=0$ only when the anomalous contribution to the $\eta'$ mass vanishes~\cite{Guoetal}, formally achieved at $N_c\rightarrow\infty$~\cite{wittenetal}. 
Thus, it is natural to expect $\theta_P\rightarrow \theta_P^{id}$ in the temperature regime where $U(1)_A$ is restored.  
This is consistent with the experimentally observed reduction of the $\eta'$ mass at finite $T$~\cite{Csorgo:2009pa} and with recent model analyses showing asymptotic ideal mixing at finite $T$~\cite{Ishii:2016dln,Lenaghan:2000ey}.

Ideal mixing is actually  an additional consequence of our present  analysis. From~\eqref{mixcond},  we get

$$\theta_P=\theta_P^{id}\Leftrightarrow  P_{ls}=0, \ P_{ll}-2P_{ss}\neq 0.$$

Therefore, our results~\eqref{wipls},~\eqref{lscond} support ideal mixing at $O(4)\times U(1)_A$ restoration. Note that $U(1)_A$ degenerates scalar and pseudoscalar partners through~\eqref{ua1deg}, so the mixing angle in the scalar sector $\sigma_{l,s}$ degenerates with $\theta_P$ in that regime.

Another limit that deserves some comments is $\theta_P=0$. From~\eqref{mixcond}, 

$$\theta_P=0\Leftrightarrow P_{ll}-2P_{ss}=P_{ls}\neq 0.$$ 

Let us now consider the correlator combination appearing in the above equation from the point of view of the WI considered in this work. 
Note that:

\begin{align}
\chi_P^{ll}-2\chi_P^{ss}-\chi_P^{ls}=-\frac{1}{\hat m}\condl 
+\frac{2}{m_s}\conds 
+2\frac{(\hat m - m_s)(\hat m + 2m_s)}{m_s^2}\chi_{5,disc},
\label{vantheta}
\end{align}
where we have used~\eqref{wichipetal},~\eqref{wichipss} and~\eqref{wipls}. 
It is plausible that~\eqref{vantheta} reaches small values around chiral restoration, 
since $-\condl(T)$ and $\chi_{5,disc}(T)$ decrease and  $-\conds (T)$ smoothly increases. 
Actually, we plot this combination in Fig.~\ref{fig:latticenonet}c using again the lattice data in~\cite{Buchoff:2013nra}. 
The neat separation between $O(4)$ and $O(4)\times U(1)_A$ found in that work  guarantees $\chi_P^{ls}\neq 0$ around chiral restoration and hence a vanishing mixing angle regime. 
We see in Fig.~\ref{fig:latticenonet}c that there is actually a $\theta_P(T)\sim 0$ region close to chiral restoration, where the combination~\eqref{wipls} develops a minimum. 
For higher $T$, $\theta_P$ moves from zero to $\theta_P^{id}$ asymptotically.
Note also that~\eqref{vantheta} vanishes in the $SU(3)$ limit, i.e. $m_s\to\hat m$ and $\conds\to \condl/2$, consistently with $\theta_P\rightarrow 0$ for $m_K=m_\pi$ at $T=0$~\cite{Guoetal}. 

Moreover, in the intermediate region between $O(4)$ and $O(4)\times U(1)_A$ restoration, 
if  both the combination in~\eqref{vantheta} and $\chi_P^{ls}$ happen to remain small, there would be  an additional sign of partner degeneracy, 
namely $2P_{ss}\sim S_{\delta\delta}$ and $2S_{ss}\sim P_{\pi\pi}$. These two identities are tested for the same lattice data in Fig.~\ref{fig:latticenonet}d, where they actually tend to  degenerate. However, if the susceptibility combination in Fig.~\ref{fig:latticenonet}c would keep on growing for higher $T$, the degeneracy in Fig.\ref{fig:latticenonet}d would not be maintained.

\subsection{Comments on the WI relating two and three point functions}
\label{sec:2p3p}

The  WI~\eqref{wipisigmaop}-\eqref{wiKKappaopS} provide constraints on specific model  and lattice analyses of partner degeneracy.
 Thus, they connect  combinations of two-point functions (correlators and susceptibilities) corresponding to degenerate partners at $O(4)$ and $U(1)_A$ restoration in the l.h.s, with three-point functions (vertices) in the r.h.s.   Actually, as they are written, the  l.h.s of~\eqref{wipisigmaop}-\eqref{wiKKappaop} should vanish at exact $O(4)$ restoration, while the l.h.s of~\eqref{wipideltaop}-\eqref{wiKKappaopS} should vanish at exact $U(1)_A$ restoration, according to our analysis in sections~\ref{sec:I01} and \ref{sec:I1/2}.  The analysis of the r.h.s of those equations would be then an additional tool to test the different partner and patterns discussed here in future lattice and model analysis.  In turn, note that assuming the pion bilinear normalization discussed in~\ref{sec:latt}, 
the r.h.s of~\eqref{wipisigmaop}-\eqref{wipideltaop}  would vanish at strict $O(4)$ restoration, which supports our results in the previous sections. 

Just for illustration, let us also mention that the r.h.s of those equations could be related to meson scattering processes. Although a meson Lagrangian relies on a low-energy description and hence may not be suitable to study the transition region, it provides a rigorous way to parametrize meson  interactions, which we will shortly employ here to emphasize the possible role of scattering processes in the meson realization of the WI. A more detailed analysis of those WI is left for future work.  
 In particular, the coupling of the $\sigma_{l,s}$ bilinears to an external scalar source in QCD is expressed in a meson Lagrangian into the $\pi\pi$, $\bar KK$ and $\eta\eta$ channels~\cite{Gasser:1984gg}.  
Therefore, the r.h.s. of identities~\eqref{wipisigmaop} and~\eqref{wiSls} are directly related 
to $\pi\pi\rightarrow\pi\pi$, $\bar KK\rightarrow\pi\pi$ and $\eta\eta\rightarrow\pi\pi$ scattering, where the $f_0(500)$ is generated.  
Actually, the role of this resonance for $O(4)$ restoration has been recently emphasized in~\cite{Nicola:2013vma,GomezNicola:2016csg}.  Similarly, the r.h.s. of~\eqref{wideltaetaop}-\eqref{wiPls} and~\eqref{wiKKappaop} connect with the $a_0(980)$ and $\kappa(800)$ resonances produced in 
$\pi\eta(\bar KK)\rightarrow\pi\eta$ and $\pi K(\pi\eta)\rightarrow \pi K$ scattering, respectively.
The r.h.s of~\eqref{wipideltaop}-\eqref{wiKKappaopS} include the effect of the $\eta'$, which couples through $A(x)$ to the $U(3)$ formulation of the chiral Lagrangian~\cite{wittenetal}.
For instance~\eqref{wipideltaop} can be expressed in terms of $\pi\eta(\eta^\prime)\rightarrow\pi\eta(\eta^\prime)$ and $\bar K K\rightarrow \pi\eta(\eta^\prime)$ processes, all in the $a_0(980)$ channel. 
Note that the light chiral limit $\hat m\rightarrow 0^+$ of the r.h.s. of equations~\eqref{wipisigmaop}-\eqref{wiKKappaopS} is in general nontrivial. 
For instance, at $T=0$ $\chi_P^\pi=\Od\left(\hat m^{-1}\right)$ and $\chi_{S}^l=\Od\left(\log\hat m\right)$~\cite{Smilga:1995qf,GomezNicolaetal}, so the r.h.s. of~\eqref{wipisigmaop} should scale at least as $1/\hat m$ at zero temperature. 

\section{Conclusions}
\label{sec:conc}

In this work, we have performed an analysis based on Ward Identities of several quantities relevant for the understanding of the  pattern of chiral symmetry restoration and the degeneration of the corresponding partners under $O(4)$ and $U(1)_A$ symmetries. In particular, our results lead to the vanishing of $\chi_{5,disc}$ and the topological susceptibility $\chi_{top}$ in the region where $O(4)$ partners are degenerated, pointing out for a $O(4)\times U(1)_A$ restoration pattern. 
This is a statement formally valid when $O(4)$ restoration is exact and hence, approximately valid depending on the strength of degeneration of chiral partners. 
In the physical case, massive light quarks and strange-quark mass contributions ($N_f=2+1$) may distort numerically this conclusion.  Our results relate in a nontrivial way chiral and $U(1)_A$ restoring quantities,  
both understood in the sense of partner degeneration, despite the difficulty to measure properly $U(1)_A$ correlators. 
In connection with this analysis, we have checked, using $N_f=2+1$ lattice results, that the thermal scaling of $\chi_{5,disc}$ and the subtracted quark condensate are close to one another, consistently with previous analysis.

In addition, for exact chiral restoration, our WI also predict $K-\kappa$ degeneracy, which in the physical case is directly linked to the subtracted lattice quark condensate. 
WI also imply additional $U(1)_A$ partner degeneracy for the $ss$ and $ls$ sectors, the former being confirmed also using lattice data.  
Regarding the $\eta-\eta^\prime$ mixing angle, our analysis is consistent with ideal  mixing at the $O(4)\times U(1)_A$ transition.  
On the massive case, a vanishing pseudoscalar mixing is expected if a sizable transient regime between $O(4)$ and $O(4)\times U(1)_A$ restoration takes place.
 
 All these conclusions have been achieved by identifying relevant combinations of correlators from those WI and studying in detail their symmetry transformation properties, which allows one for a model-independent analysis. 
In addition, through additional new WI, we have provided  useful results, testable in lattice simulations and in model analyses,  connecting partner degeneracy with specific meson vertices and processes.

\section*{Acknowledgments}
We thank Z. H. Guo, L.~Tunstall, S. Sharma and O. Phillipsen for helpful discussions.  A.G.N is very grateful to the AEC and the ITP in Bern for their hospitality and financial support. Work partially supported by  research contracts FPA2014-53375-C2-2-P, FIS2014-57026-REDT and FPA2016-75654-C2-2-P (spanish ``Ministerio de Econom\'{\i}a y Competitividad"), the DFG (SFB/TR 16, ``Subnuclear Structure of Matter'') and the Swiss National Science Foundation.


\begin{thebibliography}{99}


\bibitem{Pisarski:1983ms} 
  R.~D.~Pisarski and F.~Wilczek,
  Phys.\ Rev.\ D {\bf 29}, 338 (1984).

 \bibitem{Smilga:1995qf}
  A.~V.~Smilga and J.~J.~M.~Verbaarschot,
  Phys.\ Rev.\  D {\bf 54}, 1087 (1996).
  
  


\bibitem{Aoki:2009sc} 
  Y.~Aoki, S.~Borsanyi, S.~Durr, Z.~Fodor, S.~D.~Katz, S.~Krieg and K.~K.~Szabo,
  JHEP {\bf 0906}, 088 (2009).

\bibitem{Bazavov:2011nk} 
  A.~Bazavov {\it et al.},
  Phys.\ Rev.\ D {\bf 85}, 054503 (2012).
  
  
\bibitem{Buchoff:2013nra} 
  M.~I.~Buchoff {\it et al.},
  Phys.\ Rev.\ D {\bf 89}, no. 5, 054514 (2014).
  
  \bibitem{Ejiri:2009ac}
  S.~Ejiri {\it et al},
  Phys.\ Rev.\  D {\bf 80}, 094505 (2009).

  
  \bibitem{Gross:1980br} 
  D.~J.~Gross, R.~D.~Pisarski and L.~G.~Yaffe,
  Rev.\ Mod.\ Phys.\  {\bf 53}, 43 (1981).
 
    
    \bibitem{Eser:2015pka} 
  J.~Eser, M.~Grahl and D.~H.~Rischke,
  Phys.\ Rev.\ D {\bf 92}, 096008 (2015).

    
  \bibitem{Pelissetto:2013hqa} 
    A.~Pelissetto and E.~Vicari,    
    Phys.\ Rev.\ D {\bf 88}, 105018 (2013).
    
    \bibitem{Mitter:2013fxa} 
  M.~Mitter and B.~J.~Schaefer,
  Phys.\ Rev.\ D {\bf 89}, 054027 (2014).


\bibitem{Kapusta:2006pm} 
  J.~I.~Kapusta and C.~Gale,
  ``Finite-temperature field theory: Principles and applications'', 
  Cambridge University Press (2006). 

\bibitem{wittenetal}
  E.~Witten,
  Nucl.\ Phys.\ B {\bf 156}, 269 (1979).
  P.~Di Vecchia and G.~Veneziano,
  Nucl.\ Phys.\ B {\bf 171}, 253 (1980);
 C.~Rosenzweig, J.~Schechter and C.~G.~Trahern,
  Phys.\ Rev.\ D {\bf 21}, 3388 (1980).
  P.~Herrera-Siklody, J.~I.~Latorre, P.~Pascual and J.~Taron,
  Nucl.\ Phys.\ B {\bf 497}, 345 (1997);
  R.~Kaiser and H.~Leutwyler,
  Eur.\ Phys.\ J.\ C {\bf 17}, 623 (2000).
%

\bibitem{Csorgo:2009pa} 
  T.~Csorgo, R.~Vertesi and J.~Sziklai,
  Phys.\ Rev.\ Lett.\  {\bf 105}, 182301 (2010).

\bibitem{Kapusta:1995ww} 
  J.~I.~Kapusta, D.~Kharzeev and L.~D.~McLerran,
  Phys.\ Rev.\ D {\bf 53}, 5028 (1996).


\bibitem{Jido:2011pq} 
  D.~Jido, H.~Nagahiro and S.~Hirenzaki,
  Phys.\ Rev.\ C {\bf 85}, 032201 (2012).

\bibitem{Shuryak:1993ee} 
  E.~V.~Shuryak,
  Comments Nucl.\ Part.\ Phys.\  {\bf 21}, no. 4, 235 (1994).


\bibitem{Cohen:1996ng} 
  T.~D.~Cohen,
  Phys.\ Rev.\ D {\bf 54}, R1867 (1996).

\bibitem{Lee:1996zy} 
  S.~H.~Lee and T.~Hatsuda,
  Phys.\ Rev.\ D {\bf 54}, R1871 (1996).

\bibitem{Meggiolaro:2013swa} 
E.~Meggiolaro and A.~Morda,
Phys.\ Rev.\ D {\bf 88}, 096010 (2013).

\bibitem{Fejos:2015xca} 
  G.~Fejos,
  Phys.\ Rev.\ D {\bf 92}, 036011 (2015).


\bibitem{Heller:2015box} 
M.~Heller and M.~Mitter,
Phys.\ Rev.\ D {\bf 94}, 074002 (2016).


\bibitem{Ishii:2015ira} 
M.~Ishii, K.~Yonemura, J.~Takahashi, H.~Kouno and M.~Yahiro,
Phys.\ Rev.\ D {\bf 93}, 016002 (2016).

  
  \bibitem{Ishii:2016dln} 
  M.~Ishii, H.~Kouno and M.~Yahiro,
  Phys.\ Rev.\ D {\bf 95}, no. 11, 114022 (2017).

\bibitem{Cossu:2013uua} 
  G.~Cossu, S.~Aoki, H.~Fukaya, S.~Hashimoto, T.~Kaneko, H.~Matsufuru and J.~I.~Noaki,
  Phys.\ Rev.\ D {\bf 87}, no. 11, 114514 (2013)
  Erratum: [Phys.\ Rev.\ D {\bf 88}, no. 1, 019901 (2013)].

\bibitem{Tomiya:2016jwr} 
  A.~Tomiya, G.~Cossu, S.~Aoki, H.~Fukaya, S.~Hashimoto, T.~Kaneko and J.~Noaki,
  Phys.\ Rev.\ D {\bf 96}, no. 3, 034509 (2017).
  
  \bibitem{Aoki:2012yj}
  S.~Aoki, H.~Fukaya and Y.~Taniguchi,
  Phys.\ Rev.\ D {\bf 86} (2012) 114512.

\bibitem{Brandt:2016daq} 
  B.~B.~Brandt, A.~Francis, H.~B.~Meyer, O.~Philipsen, D.~Robaina and H.~Wittig,
  JHEP {\bf 1612}, 158 (2016).
    
   \bibitem{Aarts} 
  G.~Aarts,  C.~Allton, S.~Hands, B.~J\"ager, C.~Praki and J.~I.~Skullerud,
  Phys.\ Rev.\ D {\bf 92}, 014503 (2015). 
  G.~Aarts, C.~Allton, D.~De Boni, S.~Hands, B.~Jäger, C.~Praki and J.~I.~Skullerud,
JHEP {\bf 1706}, 034 (2017).

  
  
   \bibitem{Borsanyi:2016ksw} 
  S.~Borsanyi {\it et al.},
  Nature {\bf 539}, no. 7627, 69 (2016).

\bibitem{Petreczky:2016vrs} 
  P.~Petreczky, H.~P.~Schadler and S.~Sharma,
  Phys.\ Lett.\ B {\bf 762}, 498 (2016).


  
  
\bibitem{Nicola:2016jlj} 
  A.~G\'omez Nicola and J.~Ruiz de Elvira,
  JHEP {\bf 1603}, 186 (2016).
  
  
  \bibitem{GomezNicolaetal} 
  A.~G\'omez Nicola, J.~R.~Pel\'aez and J.~Ruiz de Elvira,
  Phys.\ Rev.\ D {\bf 82}, 074012 (2010),
  Phys.\ Rev.\ D {\bf 87}, 016001 (2013).
  
  \bibitem{Bochicchio:1985xa}
  M.~Bochicchio, L.~Maiani, G.~Martinelli, G.~C.~Rossi and M.~Testa,
  Nucl.\ Phys.\ B {\bf 262}, 331 (1985).


\bibitem{Boucaud:2009kv}
  P.~Boucaud {\it et al}, J.~P.~Leroy, A.~L.~Yaouanc, J.~Micheli, O.~Pene and J.~Rodriguez-Quintero,
    Phys.\ Rev.\ D {\bf 81}, 094504 (2010).
 
 
    \bibitem{Cheng:2010fe}
  M.~Cheng {\it et al},
  Eur.\ Phys.\ J.\ C {\bf 71}, 1564 (2011).

  
 \bibitem{Nicola:2013vma}
  A.~G\'omez Nicola, J.~Ruiz de Elvira and R.~Torres Andres,
  Phys.\ Rev.\ D {\bf 88}, 076007 (2013).
        
        
             
\bibitem{GomezNicola:2016csg} 
  A.~G\'omez Nicola, S.~Cortes, J.~Morales, J.~Ruiz de Elvira and R.~Torres Andres,
  EPJ Web Conf.\  {\bf 137}, 07016 (2017)
  [arXiv:1611.07377 [hep-ph]].


  
 

\bibitem{Guoetal} 
  Z.~H.~Guo and J.~A.~Oller,
  Phys.\ Rev.\ D {\bf 84}, 034005 (2011);
  Z.~H.~Guo, J.~A.~Oller and J.~Ruiz de Elvira,
  Phys.\ Lett.\ B {\bf 712}, 407 (2012);
  Phys.\ Rev.\ D {\bf 86}, 054006 (2012).


\bibitem{Azcoiti:2016zbi} 
  V.~Azcoiti,
  Phys.\ Rev.\ D {\bf 94}, no. 9, 094505 (2016).


\bibitem{Nicola:2011gq} 
  A.~G\'omez Nicola and R.~Torres Andres,
  Phys.\ Rev.\ D {\bf 83}, 076005 (2011).


  
 
    \bibitem{Lenaghan:2000ey} 
  J.~T.~Lenaghan, D.~H.~Rischke and J.~Schaffner-Bielich,
  Phys.\ Rev.\ D {\bf 62}, 085008 (2000).
  
  \bibitem{Gasser:1984gg}
  J.~Gasser and H.~Leutwyler,
  Nucl.\ Phys.\  B {\bf 250}, 465 (1985).

  
\end{thebibliography}
\end{document}